\documentclass[twocolumn,aps,prx,superscriptaddress,footinbib,floatfix]{revtex4-2}
\usepackage[utf8]{inputenc}

\usepackage{amsmath,amssymb,bm}
\usepackage{hyperref,xcolor,graphicx,siunitx}

\usepackage{tikzsymbols}
\usepackage{slashed}
\usepackage{ulem}

\begin{document}
\title{Spinon pairing induced by chiral in-plane exchange and the stabilization of odd-spin Chern number spin liquid in twisted MoTe$_2$}
\author{Valentin Cr\'epel}
\affiliation{Center for Computational Quantum Physics, Flatiron Institute, New York, New York 10010, USA}
\author{Andy Millis}
\affiliation{Center for Computational Quantum Physics, Flatiron Institute, New York, New York 10010, USA}
\affiliation{Department of Physics, Columbia University, New York, NY 10027, USA}

\begin{abstract}
The unusual structure and symmetry of low-energy states in twisted transition metal dichalcogenides leads to large in-plane spin-exchange interactions between spin-valley locked holes. 
We demonstrate that this exchange interaction can stabilize a gapped spin-liquid phase with a quantized spin-Chern number of three when the twist angle is sufficiently small and the system lies in a Mott insulating phase. The gapped spin liquid may be understood as arising from spinon pairing in the DIII Altland-Zirnbauer symmetry class. Applying an out of plane electric field or increasing the twist angle is shown to drive a transition respectively to an anomalous Hall insulator or an in-plane antiferromagnet. 
Recent experiments indicate that a spin-Chern number three phase occurs in twisted MoTe$_2$ at small twist angles with a transition to a quantum anomalous Hall phase as the twist angle is increased above a critical value of about $2.5^\circ$ in absence of applied electric field. 
\end{abstract}

\maketitle

\paragraph*{Introduction ---} Fractionalization of elementary particles is a fascinating hallmark of strongly correlated quantum systems. Prototypical examples are the fractional quantum Hall phases observed in sufficient clean two dimensional semiconductors at high magnetic fields~\cite{stormer1999fractional}, which host quasiparticles and chiral edge modes carrying fractions of the electronic charge, as manifested in a fractionally quantized Hall conductivity. Analogous fractional Chern insulator (FCI) phenomena have recently been observed at zero applied field in MoTe$_2$ homobilayers with a relative twist of $3.4-4.0^\circ$. The FCI behavior occurs in ferromagnetic states with a  spontaneously broken time-reversal symmetry~\cite{cai2023signatures,xu2023observation,zeng2023thermodynamic,park2023observation,redekop2024direct}. These findings highlight the interest of twisted transition metal dichalcogenides (TMDs) as exceptional platforms intertwining topological and correlation effects and enabling the systematic study of novel phases of matter.

Fractionalization of electronic degrees of freedom into Majorana non-abelian quasiparticles has also been predicted to occur in spin systems featuring non-isotropic spin exchange interactions~\cite{kitaev2006anyons} and in chiral $p$-wave superconductors~\cite{read2000paired}. In fact, any time-reversal invariant superconducting state in the DIII Altland-Zirnbauer class~\cite{altland1997nonstandard,ryu2010topological} is characterized by an integer topological bulk invariant $C_s$ counting the number of helical Majorana modes at their edges~\cite{leijnse2012introduction,crepel2023topological}. When odd, $C_s$ signals existence of Majorana quasiparticles explicit in a fractional quantum spin-Hall conductivity $C_s e^2/ (2 h)$.

Recently, non-local edge transport data consistent with a fractional quantum spin-Hall effect with $C_s = 3$ has been reported in a $2.1^\circ$-twisted MoTe$_2$ homobilayer at the filling of $\nu=-3$ holes per moir\'e unit cell~\cite{kang2024observation}. As far as is known, this state does not exhibit a spin polarization, unlike the FCI states observed at larger twist angles. Theory suggests that the ferromagnetic FCI states with charged edge modes observed for twists $\geq 3^\circ$ are stabilized by repulsive interactions between carriers~\cite{li2021spontaneous,crepel2023anomalous}, but the observation of the insulating spin-Hall state with odd $C_s$ apparently requires pairing between either the original carriers or the quasi-particles of the system~\cite{kang2024observation}.

Here, we present a theory for the odd spin-Chern number phase in $\nu=-3$ small-angle twisted TMDs and its twist angle or displacement field-drive transitions to other better-known phases. The theory is based on three crucial observations. First, the typical kinetic energy on the moir\'e scale $E_{\rm kin} = \frac{\hbar^2 \theta^2}{2m^* a^2}$, with $m^*$ and $a$ the effective mass and lattice constant of the TMD, decreases faster than interactions as the twist angle is reduced. At small angles, the states at integer fillings must be Mott insulators, with charge degrees of freedom frozen and a low-energy physics governed by an effective spin model. Second, we show that this spin model is classically frustrated and features non-isotropic exchange interactions arising from the strong spin-orbit coupling of TMDs; two favorable ingredients for the emergence of spin-liquid phases~\cite{balents2010spin,kim2024theory,divic2024chiral}. 
Finally, we show using exact-diagonalization on finite clusters and a spinon mean-field analysis that the dominant spin-interaction of the spin model -- a chiral in-plane exchange -- mediates attractive interactions between spinons that stabilize a spinon superconductor characterized by an odd total spin-Chern number $C_s = |\nu| = 3$. The predicted gap and stability range as a function of twist angle of this paired spinon phase agrees with experiments. For larger angles, \textit{i.e.} above the bandwidth-tuned Mott transition, our mean-field calculations predict a transition to a $C=1$ Chern insulator that has very recently been experimentally reported~\cite{park2024ferromagnetism,xu2024interplay}, while as interlayer electric (``displacement'') field is increased from zero we predict a transition to an in-plane antiferromagnet.

\begin{figure}
\centering
\includegraphics[width=\columnwidth]{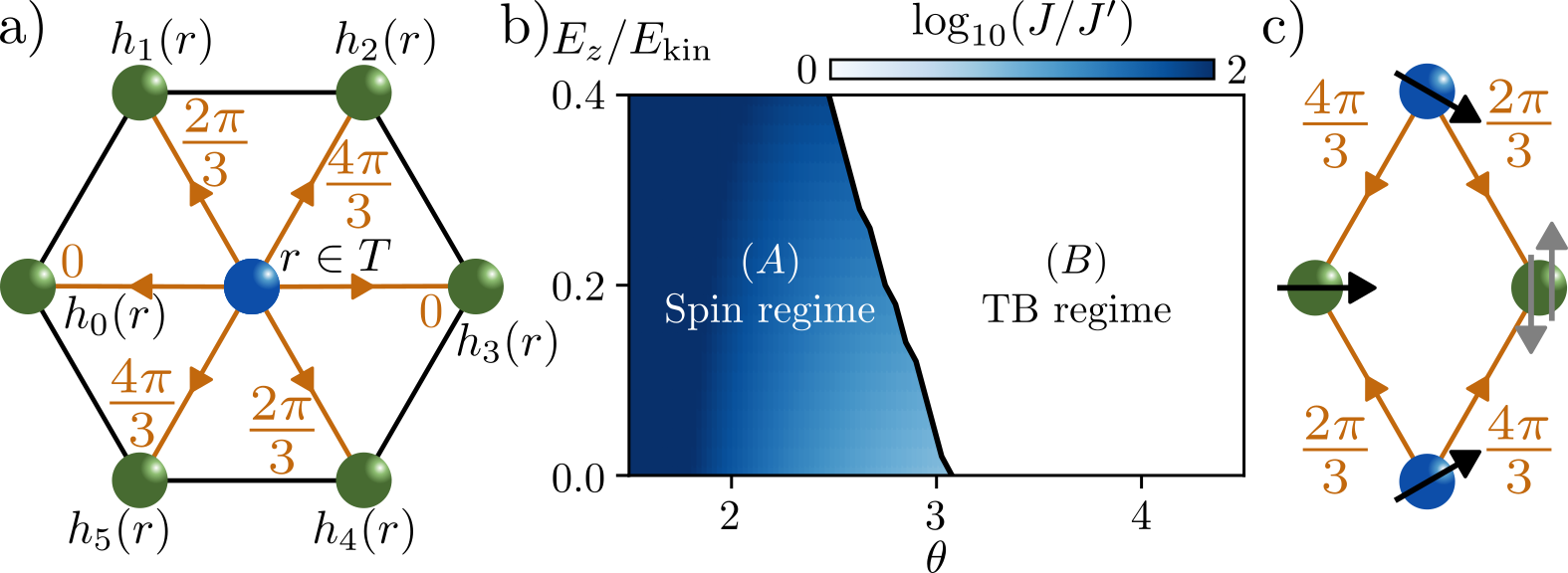}
\caption{a) Sketch of the lattice with triangular (blue) and honeycomb sites (green). The U(1) chiral spin exhange $J$ and the SU(2)-symmetric Heisenberg interaction $J$ of Eq.~\ref{fig_tworegimes} are respectively depicted with orange arrows and black lines between neighboring sites of the lattice.  
b) Qualitative metal-insulator phase diagram, where the Mott transition is determined as the $E_z$ where the bandwidth of Eq.~\ref{eq_tb} equals $\max (U_{\rm T/H})$ (black line) using the realistic values of the parameters obtained in Ref.~\cite{crepel2024}. In regime ($A$) where  the charge degree of freedom is frozen and the system is governed by the effective spin model Eq.~\ref{eq_spinmodel}. The color density in regime (A) shows the ratio $J/J'$. 
c) The chiral spin-exchange is classically frustrated on all elementary parallelograms involving two triangular sites and their common honeycomb neighbors. 
}   
\label{fig_tworegimes}
\end{figure}

\paragraph*{Model for $\nu=-3$ TMD homobilayers ---} We~\cite{crepel2024} and others~\cite{qiu2023interaction,xu2024maximally} have recently derived a symmetry-adapted tight-binding Hamiltonian capturing the low-energy properties of twisted TMD homobilayers. 
To faithfully capture the Chern number of the topmost two valence bands of the bilayer, which can be equal and non zero in certain regime of twist angle~\cite{zhang2024polarization}, the Hamiltonian features three bands for each spin-valley component $\tau=\pm$ originating from the three orbitals depicted in Fig.~\ref{fig_tworegimes}a. 
The first one (T) lies at the triangular site (1a Wyckoff position), has equal weight in the two layers and possesses a $p_\tau$ orbital character; the two others (H) are  located at honeycomb sites (1b/1c Wyckoff positions) and have opposite layer polarization and possess an $s$ wave orbital character. Tunneling and spin-exchange terms between the T and H-centered orbitals acquire phases due to the $p_\pm$-wave nature of the T orbitals, leading to an interacting tight-binding Hamiltonian $H_{\rm tb} = H_{\rm tun}^+ + H_{\rm tun}^- + H_{\rm pot} + H_{\rm int}$ where
\begin{widetext}
\begin{align} \label{eq_tb}
H_{\rm tun}^\tau & = t' \sum_{\langle r,r'\rangle_{\rm H}} (c_{r',\sigma}^\dagger c_{r,\sigma} +hc) +  t \sum_{r\in {\rm T}} \sum_{n=0}^5 \left( e^{\frac{i\tau n \pi}{3}} c_{h_n(r),\sigma}^\dagger c_{r,\sigma} +hc \right) , \quad H_{\rm pot}  =  \sum_{r\in {\rm T}, \sigma} \delta n_{r,\sigma}+\sum_n \frac{E_z}{2} (-1)^n n_{h_n(r),\sigma} , \notag \\
H_{\rm int} & = U_T \sum_{r\in {\rm T}} n_{r,\uparrow} n_{r,\downarrow} + U_H \sum_{r\in {\rm H}} n_{r,\uparrow} n_{r,\downarrow}  +\frac{1}{2} \sum_{r\in {\rm T}} \sum_{n=0}^5 V n_{h_n(r)} n_{r} + J \left[ e^{\frac{2 i n \pi}{3}} c_{h_n(r),\uparrow}^\dagger c_{h_n(r),\downarrow} c_{r,\downarrow}^\dagger c_{r,\uparrow} + hc \right] , 
\end{align}
\end{widetext}
with $\delta>0$ the energy potential difference between T and H sites, $E_z$ an out-of-plane electric field, $\sigma = \uparrow/\downarrow$ the spin component locked to the valley index $\tau=+/-$, $\langle r,r'\rangle_{\rm H}$ running over nearest neighbors in H, and $h_n(r)$ with $n=0,\cdots,5$ the honeycomb neighbors of the triangular site $r \in {\rm T}$ counted from the leftmost one in clockwise order (see Fig.~\ref{fig_tworegimes}a). 
We  have only retained the tunneling of smallest spatial extent $t,t'$, which scale as $\theta^2$ in the small angle limit $\theta \to 0$ due to the quadratic dispersion of the monolayers. The dominant interaction is an on-site repulsion $U$ (which may have somewhat different magnitude on the two inequivalent sites because of different spatial spread of the relevant Wannier functions.
We have also included the next most dominant interaction coefficients, a direct Coulomb term $V$ and an intersite exchange $J$. 
They both originate from projecting the bare Coulomb repulsion onto the Wannier functions of the tight-binding model, and remain relatively strong compare to other non-local interactions due to the tails of the T Wannier orbitals that directly overlap with the sharply localized Wannier functions at neighboring honeycomb sites. Note that the orbital symmetry implies that the exchange is \textit{chiral}.

Since $t,t'$ decrease more rapidly with twist angle than do the interaction parameters, we can distinguish a correlation-induced insulator at small twist angle ($A$) and a metallic regime ($B$) where the overall single-particle bandwidth is larger than the interaction scales. The two regimes are separated by a Mott transition, whose location is sketched in Fig.~\ref{fig_tworegimes}b as the position where the total (three-band) hole-bandwidth equals $\max(U_{\rm T/H})$ using realistic values for the parameters obtained in Ref.~\cite{wang2023fractional,crepel2024} as a function of $\theta$ and $E_z$. While microscopic details not included in our theory, including in-plane lattice relaxations and further neighbor hoppings, likely shift the precise location of the Mott transition, the existence of the two phases, the approximate value of twist angle where the Mott transition occurs, and the qualitative displacement field dependence of the phase boundary are expected to be robust. 
In the rest of the paper, we study the phases arising in ($A$) using an effective spin model and a spinon mean-field theory, and describe the neighboring phases in region ($B$) by solving Eq.~\ref{eq_tb} within the Hartree-Fock approximation.

At the filling of three holes per moir\'e unit cell, the spin Hamiltonian $H_{\rm sp}$ governing the low-energy physics in regime ($A$) is obtained by placing one hole at each site of the lattice, thereby avoiding the strong energy costs $U_{\rm T/H}$. Keeping only the dominant spin-interactions terms between nearest neighbors, we obtain 
\begin{equation}\label{eq_spinmodel}
H_{\rm sp} = J' \sum_{\langle r,r'\rangle_{\rm H}} \vec{s}_{r'} \cdot \vec{s}_{r} + J \sum_{r,n} \left[ e^{\frac{2 i n \pi}{3}} \vec{s}_{h_n(r)}^{\,+} \vec{s}_{r}^{\,-} + hc \right] ,
\end{equation}
which is sketched in Fig.~\ref{fig_tworegimes}a, and where $J' = 4 t'^2/U_{\rm H}$. This Hamiltonian only depends on the ratio $J/J'$, which we depict in Fig.~\ref{fig_tworegimes}b using the values from Ref.~\cite{crepel2024} for concreteness. 
Due to the different origin of the two spin interaction terms -- $J$ is induced by the bare Coulomb repulsion, while $J'$ arises from second order kinetic processes -- the ratio $(J/J')$ is controlled by the twist angle and decreases as $\theta$ grows to reaches its smallest value at the boundary with region ($B$) where it equals $\simeq 6$. This highlights the predominance of the bare Coulomb in-plane spin-exchange over Heisenberg interactions.

In the physically relevant regime $J \gg J'$, the model Eq.~\ref{eq_spinmodel} is geometrically frustrated because of the chiral nature of the exchange. This is most easily seen when $J'=0$ by looking at an elementary parallelogram containing two triangular sites and their two common honeycomb neighbors as sketched in Fig.~\ref{fig_tworegimes}c. The phases of the chiral exchange that do not add up to a multiple of $2\pi$ when accumulated in a loop around the parallelogram, prevent simultaneous minimization of all nearest neighbor exchange terms. 
At the quantum level, such geometric frustration can give rise to interesting spin liquid phases~\cite{balents2010spin}, which we now investigate within a mean field approach.

\paragraph*{Spinon quantized spin Hall effect at low angle ---} To analyse the effects of spin frustration we employ the standard fermionic spinon representation of the spin operators and treat the resulting Hamiltonian in mean field theory by factorizing the four spinon terms. Appendix ~\ref{app_finiteclusters} presents an exact small cluster analysis of the spinon Hamiltonian that confirms the essential conclusion of the mean field theory.  In the mean field theory we introduce all order parameters compatible with the translation, time-reversal, mirror and three-fold rotation symmetries of the twisted homobilayer~\cite{qiu2023interaction}. In the experiments of Ref.~\cite{kang2024observation}, no indications of such symmetry broken phase has been observed, which phenomenologically justifies our choice to neglect order parameters involving translation or rotational symmetry breaking.

For clarity, we introduce the various mean-field parameters in a hierarchical manner, starting with the $J'=0$ limit. Performing the standard spinon mean-field decoupling~\cite{wen1991mean} we find
\begin{align} \label{eq_spinonmeanfieldJ}
H_{\rm mf}^J = &-J \sum_{r,n} \chi e^{\frac{in\pi}{3}} (f_{h_n(r),\uparrow}^\dagger f_{r,\uparrow} + f_{r,\downarrow}^\dagger f_{h_n(r),\downarrow}) + hc \notag\\ 
& - J \sum_{r,n} \Delta^\ell_n ( f_{h_n(r)\downarrow} f_{r\uparrow} + f_{h_n(r)\uparrow}^\dagger f_{r\downarrow}^\dagger ) + hc , 
\end{align}
with spinon hopping $\chi =  \langle f_{h_{0},\uparrow}^\dagger f_{r,\uparrow} \rangle$ and pairing $\Delta^\ell_n = e^{i\ell n \pi /3} \Delta$, where $\ell$ denotes the angular momentum of the pairing function and $\Delta = \langle f_{h_{0},\downarrow} f_{r,\uparrow} \rangle$ its amplitude. Note that while the state with $\Delta\neq 0$ bears a formal similarity to a superconducting state, the U(1) charge symmetry remains unbroken because  spinons do not carry charge. We refer to the spectrum obtained from  Eq.~\ref{eq_spinonmeanfieldJ} as the ``spinon band structure'' (for $\Delta=0$) or the ``spinon Bogoliubov-deGennes'' (spinon BdG) spectrum (for $\Delta\neq 0)$.

\begin{figure}
\centering
\includegraphics[width=\columnwidth]{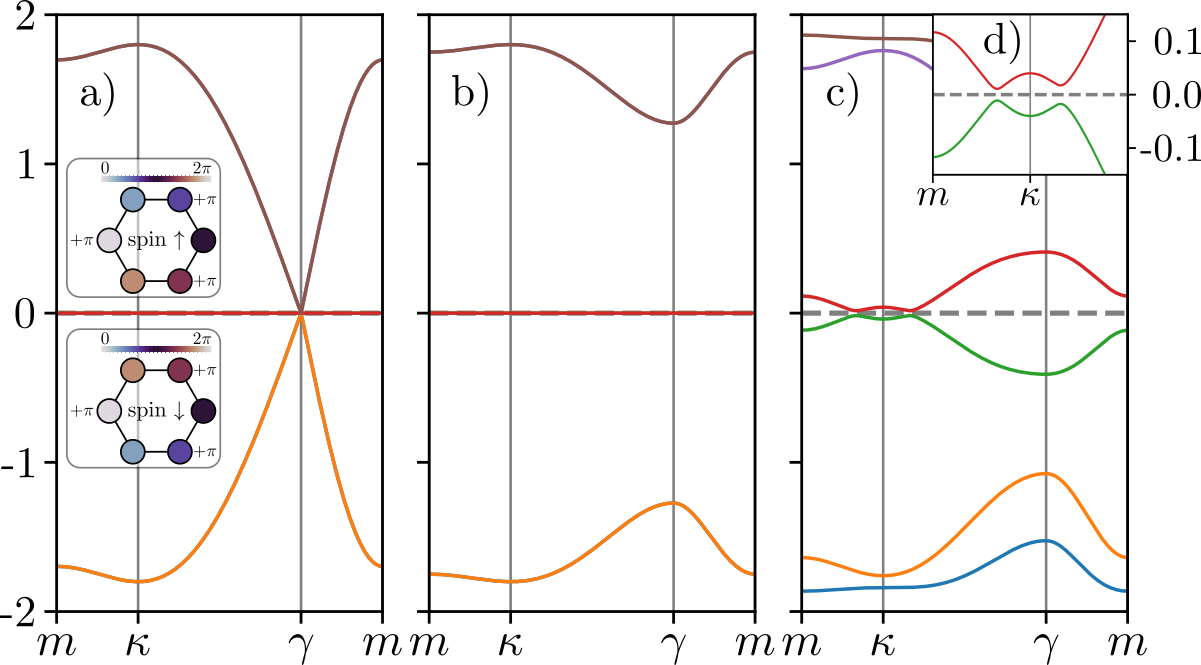}
\caption{Typical spinon BdG band structure highlighting the effect of the different mean-field parameters: (a) only includes $\chi = 0.2$, (b) adds a non zero pairing field $\Delta = 0.1$, (c) incorporates the honeycomb bond order $\chi' = 0.15$ and a non-zero chemical potential $\mu = 0.04$, and (d) finally has $\Delta' = 0.05$. All energies are given in units of $J'$ and we fixed $J = 6J'$. 
The insets in (a) depict the compact localized states spanning the spin-degenerate flat band, where the color/size represent the phase/amplitude on honeycomb sites.}
\label{fig_spinonbandsillustrative}
\end{figure}

Setting $\Delta=0$ we obtain a band structure of the form shown in Fig.~\ref{fig_spinonbandsillustrative}a. Three two-fold spin degenerate bands are evident, two dispersing and one flat in each spin sector. The flat bands express the extensive ground state degeneracy arising from the geometric frustration of the original spin model; their wavefunctions are fully localized on the honeycomb lattice sites due to the bipartite nature of the Hamiltonian, and display a chiral orbital structure, as shown in the insets of Fig.~\ref{fig_spinonbandsillustrative}a. 
The spin-degenerate dispersive bands intersect these flat bands at $\gamma$, the center of the Brillouin zone, forming for each spin a three-fold degenerate  ``spin-1'' structure analogous to that found in the Lieb lattice~\cite{lieb1989two}. 

There are three pairing states, labelled by orbital index $\ell=-1,0,1$. When $\ell=0$, the pairing potential can be viewed as a spin-orbit coupling for the spin-1 crossing, which gaps out the spinon band at $\gamma$, as shown in Fig.~\ref{fig_spinonbandsillustrative}b. The spinon Hamiltonian remaining bipartite, the two perfectly flat bands remain pinned at zero energy.  The other independent pairing channels $\ell = \pm 1$ do not gap out the crossing at $\gamma$ and are therefore not energetically favored. For the rest of this paper we focus on $\ell=0$ and do not write the $\ell$ explicitly.

The effect of the dominant interaction having been explained, we now re-introduce the SU(2) Heisenberg interactions between honeycomb sites, and similarly perform a mean-field decoupling to obtain
\begin{align} \label{eq_spinonmeanfieldJprime}
H_{\rm mf} &= H_{\rm mf}^J + \frac{J' h_z}{2} \sum_{r,n} (-1)^n [n_{h_n(r),\uparrow}^f - n_{h_n(r),\downarrow}^f ] \notag \\ &-J' \sum_{\langle r, r'\rangle_H, \sigma} \chi' 
( f_{r'\uparrow}^\dagger f_{r\uparrow} + f_{r\downarrow}^\dagger f_{r'\downarrow} ) + hc \notag\\ 
& - \frac{J'}{2} \sum_{\langle r, r'\rangle_H} \Delta' ( f_{r'\downarrow} f_{r\uparrow} + f_{r'\uparrow}^\dagger f_{r\downarrow}^\dagger ) + hc , 
\end{align}
with the bond order $\chi' = \langle f_{h_0, \uparrow}^\dagger f_{h_1, \uparrow} \rangle$ and pairing field $\Delta' = \langle f_{h_0, \downarrow} f_{h_1, \uparrow} \rangle$ defined as above. We have also introduce an out-of-plane staggered magnetization $h_z = \frac{1}{2} \langle n_{h_0,\uparrow}^f - n_{h_0,\downarrow}^f \rangle$ which is allowed by symmetry and explicitly distinguishes the sublattices of the hexagonal lattice. The inclusion of $\chi'$ and $h_z$ introduces dispersion in the flat bands and unpins the chemical potential $\mu_f$ from zero energy (Fig.~\ref{fig_spinonbandsillustrative}c), leading to intersecting BdG bands whose crossings away from high-symmetry points are finally opened by the pairing $\Delta'$  as shown in Fig.~\ref{fig_spinonbandsillustrative}d.

\begin{figure}
\centering
\includegraphics[width=\columnwidth]{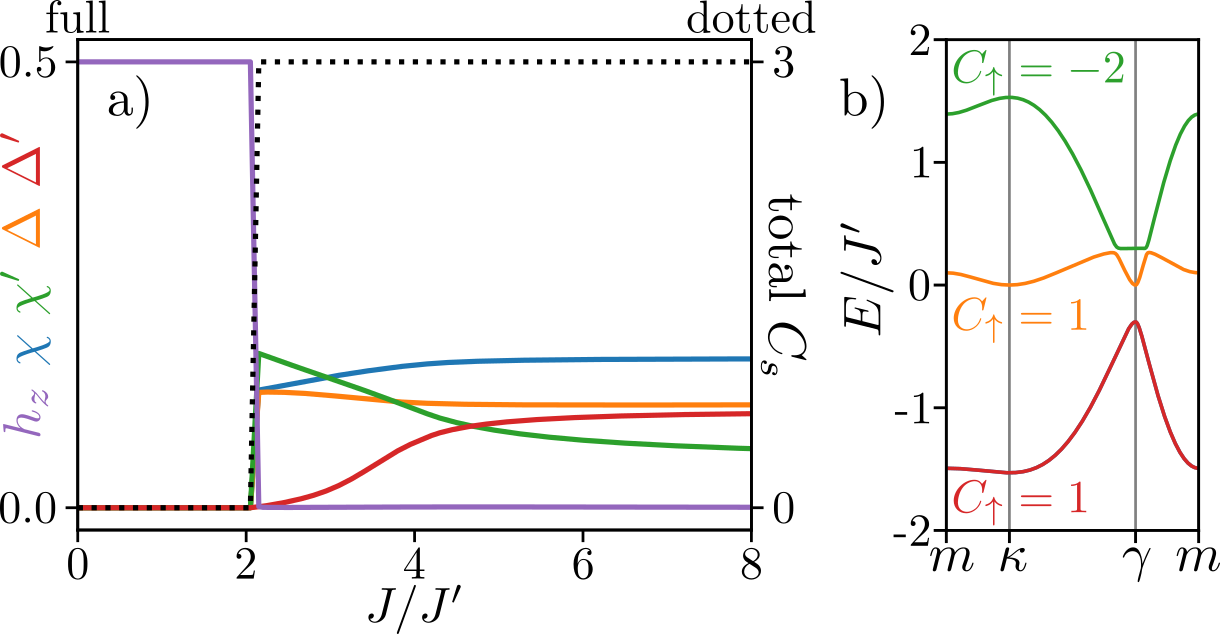}
\caption{a) Mean-field parameters of Eq.~\ref{eq_spinonmeanfieldJprime} at self-consistency (full lines, left $y$-axis), separating an antiferromagnet phase ($J<2J'$) from the spinon BdG regime ($J>2J'$) illustrated in Fig.~\ref{fig_spinonbandsillustrative}c. We also give the total Chern number at filling $\nu=-3$ (dotted line, right $y$-axis). b) Band structure of Eq.~\ref{eq_spinonmeanfieldJprime} in the ``normal'' state ($\Delta = \Delta' = 0$) for $\chi = 0.17$ and $\chi' = 0.1$ [close to the $J\gg J'$ values form (a)]. The Chern number of the spinon-up bands are indicated, those of spinon-down bands are opposite.}
\label{fig_meanfieldparameters}
\end{figure}

Now that the meaning and effect of the various mean-field parameters is clear, we present in Fig.~\ref{fig_meanfieldparameters}a the order parameter values found from  full self consistent solution of the mean field equations as a function of $J/J^\prime$ at displacement field $E_z=0$ (full lines). At small $J/J^\prime$ we find a staggered antiferromagnet ($h_z\neq0$, all other non-local order parameters vanishing; as $J$ passes through a critical value $\approx 2J^\prime$ there is a discontinuous transition to a spinon BdG  phase with $h_z=0$ and all non-local orders non-zero. The parameter estimates from microscopics discussed above suggest that throughout the Mott phase $J/J^\prime \gtrsim 6$, so that the Mott phase of t-MoTe$_2$ is expected to be described by the BdG spinon Hamiltonian Eq.~\ref{eq_spinonmeanfieldJprime} with typical dispersion depicted in Fig.~\ref{fig_spinonbandsillustrative}c.

Next, we investigate the topological properties of the spinon band structure. Because spinons do not carry charge, there is no electrical Hall conductivity associated to a spinon band with non-zero Chern number, but the non-zero spin-Chern number yields a spin-edge mode and a quantized spin-Hall conductivity. We have computed the spin-Chern number of the occupied spinon bands, using the method of Refs.~\cite{sheng2003phase,fujui2005chern,sheng2006quantum,fukui2007fukui}. The sum of these Chern numbers $C_s$ is depicted in Fig.~\ref{fig_meanfieldparameters}a (dotted line), it is equal to zero in the antiferromagnetic phase and to three in the spinon BdG phase relevant for bilayer TMDs (see Fig.~\ref{fig_tworegimes}a). 

The value of the spin-Chern in these two regimes may be understood from elementary arguments. In the antiferromagnetic phase ($J<2J'$), the mean field Hamiltonian Eq.~\ref{eq_spinonmeanfieldJprime} only has (spin-dependent) sublattice potential terms and no tunneling, \textit{i.e.} it is an atomic insulator which necessarily has vanishing Chern numbers. 
To understand the spinon BdG regime we first consider the ``normal'' state obtained by setting all pairing terms to zero $\Delta = \Delta' = 0$ so that the mean-field Hamiltonian Eq.~\ref{eq_spinonmeanfieldJprime} becomes a spin-conserving tight-binding model, whose band structure is shown in Fig.~\ref{fig_meanfieldparameters}b. The latter inherits chiral phases from the chiral in-plane exchange (see Eq.~\ref{eq_spinonmeanfieldJ}), making it analogous to the time-reversal symmetric $\pi/3$-Hofstadter model with the mild difference that $\chi \neq \chi'$. For $\chi = 0.17$ and $\chi' = 0.1$ representative of the values obtained for $J\gg J'$ in Fig.~\ref{fig_meanfieldparameters}a, this difference does not change the topological properties of the model, and bands carry spin-Chern number $(1,1,-2)$ from low to high energy. At filling $\nu=-3$, the two lowest bands are fully filled leading to a spin-Chern number of two. The two central bands carry the same spin-Chern number, but are half-filled. Upon reintroducing the pairing fields, a gap opens at the chemical potential, and these bands contribute to half of their total Chern number. Altogether, this gives rise to a total spin-Chern number $C_s = 3$. This odd integer value is rooted in the pairing gap opened at the spinon chemical potential, which emerges from to the in-plane exchange of our model (see Fig.~\ref{fig_smallclusters}). 

Finally, let us comment on the typical temperature at which the quantized spin-Hall effect with $C_s = 3$ might be observed. The gap of the mean-field band structure at self-consistency for $J = 6J'$ is equal to $G = 0.03 J'$. This gap is primarily set by $\chi'$ and $\Delta'$ (Fig.~\ref{fig_spinonbandsillustrative}) and remains of similar order in the relevant experimental regime $J \gg J'$ (Fig.~\ref{fig_tworegimes}a). Introducing the typical values $t'=\SI{7}{\milli\electronvolt}$ and $U_{\rm H}=\SI{60}{\milli\electronvolt}$~\cite{crepel2024}, we find $G \simeq \SI{0.1}{\milli\electronvolt}  \sim \SI{1}{\kelvin}$, consistent with the temperature range where the $C_s = 3$ quantized spin-Hall effect has been experimentally observed~\cite{kang2024observation}.

\begin{figure}
\centering
\includegraphics[width=\columnwidth]{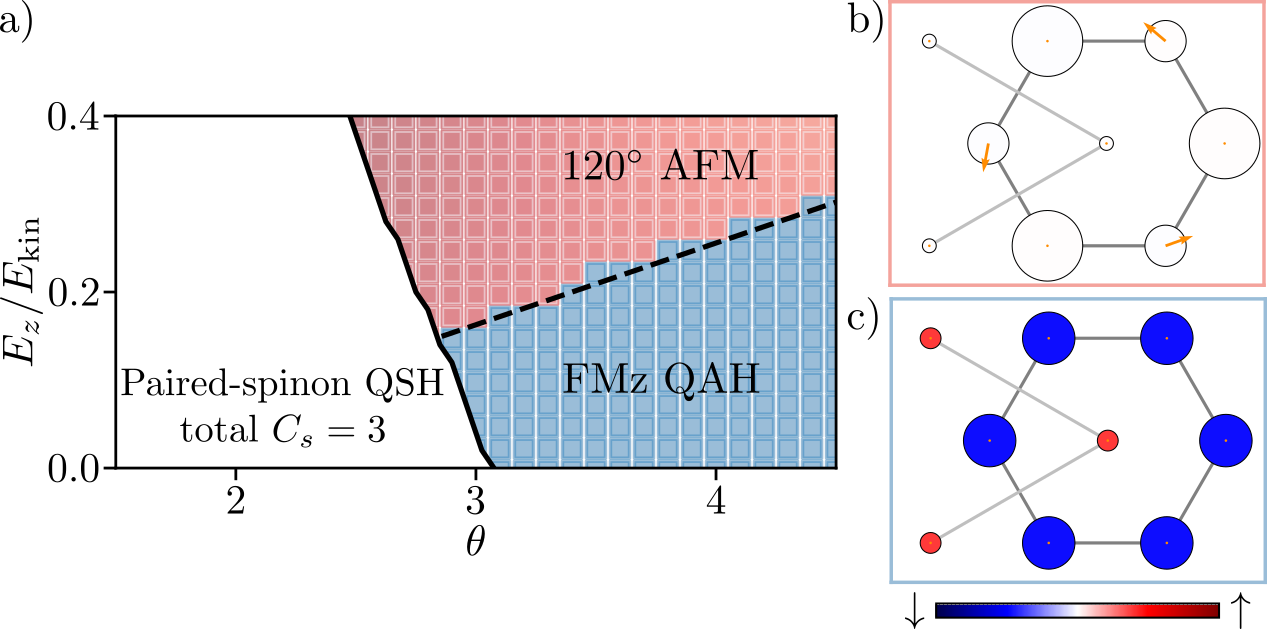}
\caption{a) Hartree-Fock phase diagram of Eq.~\ref{eq_tb} in region ($B$) of Fig.~\ref{fig_tworegimes}, see Ref.~\cite{crepel2024} for methods. We find an 120$^\circ$ N\'eel anti-ferromagnet for large $E_z$ (120$^\circ$ AFM, depicted in b), and a out-of-plane ferromagnetic insulator exhibit an quantum anomalous Hall effect (FMz QAH, depicted in c). Region ($A$) is the paired spinon phase described in Fig.~\ref{fig_spinonbandsillustrative} and~\ref{fig_meanfieldparameters}, and carries a total spin-Chern number $C_s=3$.}
\label{fig_finalphasediag}
\end{figure}

\paragraph*{Larger angle and neighboring phases ---} We now identify the phases directly neighboring the $C_s = 3$ spin-Hall phase in the tight-binding limit ($B$) of Fig.~\ref{fig_tworegimes}a by treating  Eq.~\ref{eq_tb} within the Hartree-Fock approximation following the method described in Ref.~\cite{crepel2024}. Similar calculations have been performed in many references~\cite{devakul2021magic,qiu2023interaction,zang2021hartree,kiese2022tmd,hu2021competing,fan2024orbital}, and we only briefly sketch the results here. For simplicity, we only keep $U_{\rm H/T}$ that largely dominate over the other interactions scales, and assume them equal. Our results shown in Fig.~\ref{fig_finalphasediag} show an out-of-plane ferromagnetic insulator carrying Chern number one, \textit{i.e.} a quantum anomalous Hall ferromagnet, at all angles for small applied electric field $E_z$; and a topologically trivial 120$^\circ$ N\'eel anti-ferromagnet for large displacement fields.

\paragraph*{Conclusion ---} Our results for twisted transition metal dichalcogenides at the filling of three holes per moir\'e unit cells are summarized in the phase diagram of Fig.~\ref{fig_finalphasediag}. 
Noticing that the total bandwidth of the three topmost moir\'e bands decreased with the twist angle, we were able to identify two drastically different regime (Fig.~\ref{fig_tworegimes}): 
\begin{itemize}
\item[($A$)] For small angles, the charge degree of freedom is frozen and the system must be described by an effective spin model, that feature a strong classically-frustrated chiral in-plane exchange. We have shown by exact diagonalization of small clusters (App.~\ref{app_finiteclusters}) and self-consistent mean-field (Fig.~\ref{fig_meanfieldparameters}) that this term drives pairing between spinons. The pairing gap enables the formation of an insulating state featuring an odd spin-Chern $C_s = |\nu| = 3$, which could correspond to the one observed in Ref.~\cite{kang2024observation} at the twist angle $\theta = 2.1^\circ$. This effective spin theory only applies if the three holes within each unit cell localize to form a $\sqrt{3}\times\sqrt{3}$-reduced triangular lattice. Scanning tunneling microscopy could therefore rule out or help validate our analysis. 
\item[($B$)] The large-angle regime is described by more conventional interacting tight-binding models in the original fermionic degrees of freedom and was studied within the Hartree-Fock approximations. 
In absence of applied electric field $E_z$, we find a $C=1$ ferromagnetic Chern insulator consistent with the quantized Hall conductance that has just been reported at $\nu=-3$ in samples with twist angles $\theta = 2.6-3.0^\circ$~\cite{park2024ferromagnetism,xu2024interplay} . 
\end{itemize}

Let us highlight that the transition between the paired spinon QSH phase and the time-reversal breaking QAH phase has not been studied in detailed here, as our spinon mean-field theory only applies deep in the Mott phase. The ferromagnetic Chern insulator can nevertheless be easily discriminated experimentally by the value of the Hall conductivity or by circular dichroism, which allows to experimentally determine on which side of the Mott transition (solid black line in the phase diagram of Fig.~\ref{fig_finalphasediag}a) the system lies. 
Our work calls for more precise numerical simulations able to capture the phases in regime ($A$) and ($B$) within the same framework, \textit{i.e.} capturing the Mott transition and spin physics from Eq.~\ref{eq_tb} without relying on the effective spin model Eq.~\ref{eq_spinmodel}, but also potential competing magnetic phases in region ($A$). 
Let us comment that such a method will likely require a multi-band description. Indeed, interaction-induced band mixing between the first two bands was already observed to be important at twist angle $3.7^\circ$ in twisted MoTe$_2$ to quantitatively capture the landscape of observed phases~\cite{yu2024fractional}. Band mixing is expected to be further enhanced at low angle where the single-particle band gaps and widths are further reduced~\cite{wang2024higher,xu2024multiple}.

\paragraph*{Acknowledgements ---}  V.C. thanks the entire group of K. F. Mak and J. Shan, and in particular K. Kang, for insightful discussions and for their hospitality.  AJM acknowledges support from the NSF MRSEC program through the Center for Precision-Assembled Quantum Materials (PAQM) NSF-DMR-2011738. The Flatiron Institute is a division of the Simons Foundation.

\appendix 

\section{In-plane exchange mediate spinon pairing} \label{app_finiteclusters}

\begin{figure}
\centering
\includegraphics[width=0.8\columnwidth]{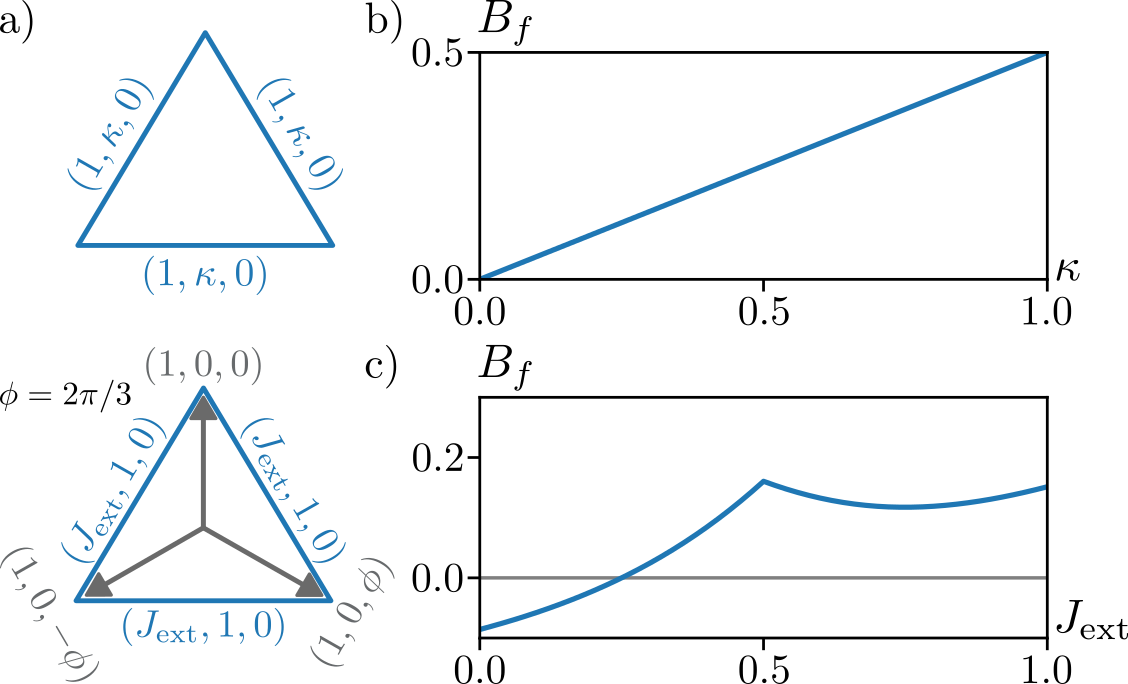}
\caption{(a) Sketch of the triangle and tripod spin clusters where the triplets $(J,\kappa,\theta)$ encode the Hamiltonian of the cluster using Eq.~\ref{eq_spinclusters}. (b) The spinon binding energy $B_f$ at half-filling for the triangle as a function of $\kappa$. (c) $B_f$ for the tripod as a function of $J_{\rm ext}$. Depending on the geometry, pure in-plane exchange interactions ($\kappa=J_{\rm ext}=0$) lead to the absence of repulsion $B_f=0$ (b), or an effective spinon attraction $B_f<0$ (c).}
\label{fig_smallclusters}
\end{figure}

To illustrate how the in-plane spin exchange can provide an effective attractive interaction between spinons, we here study models closely related to Eq.~\ref{eq_spinmodel} on small spin clusters: a triangle and a tripod shown in Fig.~\ref{fig_smallclusters}a. We find that the pure in-plane exchange spin interactions generically lead to the absence of repulsion between spinons, and in some geometries even yields an effective spinon-attraction. 

The spin clusters are ruled by the all-to-all spin Hamiltonian 
\begin{equation} \label{eq_spinclusters} \begin{split}
& H_{\rm cl} = \sum_{i<j} J_{ij} h_{ij} (\kappa_{ij} , \theta_{ij}) , \\ 
& h_{ij} (\kappa , \theta) = \kappa s_i^z s_j^z + \frac{1}{2} \left[ e^{i\theta} s_j^+ s_i^- + hc \right] , 
\end{split} \end{equation}
where the triplets $(J_{ij},\kappa_{ij},\theta_{ij})$ are depicted on the $ij$ links of the clusters shown in Fig.~\ref{fig_smallclusters}a. The SU(2) Heisenberg interaction corresponds to $\kappa=1$, while the purely in-plane exchange has $\kappa=0$. Note that the choice of $\theta$ phases for the tripod cannot be absorbed by local rotation of the spins around the $z$-axis when $J_{\rm ext} \neq 0$. The spin Hamiltonian $H_{\rm cl}$ can be formally rewritten in terms of spinon fermionic operators $f_{\uparrow/\downarrow}$ using the substitution $\Vec{s} = f_\alpha^\dagger [\Vec{\sigma}]_{\alpha\beta} f_\beta /2$ that is only exact when we further enforce $\sum_\alpha  n_\alpha^f = \sum_\alpha f_\alpha^\dagger f_\alpha= 1$ at each site. 
In the spinon theory, this half-filling constraint is relaxed and treated on average by inclusion of a chemical potential $\mu_f$.
We measure the sign of spinon interactions at half-filling by solving the quartic spinon-Hamiltonian $H$ and extracting the binding energy $B_f = E_0 - 2E_{-1} + E_{-2}$, where $E_N$ denotes the ground state energy for a system with $N$ spinon with $N$ measured from half-filling. 

For the triangle, for which all $\theta=0$ and all $\kappa$ are equal, we see that $B_f \geq 0$, which means an effective repulsion between spinons (Fig.~\ref{fig_smallclusters}b). The strength of this repulsion decreases with $\kappa$, a feature shared by all finite-size cluster that we have investigated (\textit{e.g.} a single bond). The repulsion eventually vanishes in the limit of purely in-plane exchange interactions ($\kappa=0$). Surprisingly, in some more complicated geometries, we observe that $B_f < 0$ when all the coupling are purely in-plane. This is the case for the tripod (Fig.~\ref{fig_smallclusters}c), when the SU(2) exchange in the exterior ring vanishes $J_{\rm ext} = 0$. This unveils an effective pairing potential between spinon at half-filling, which is weakened and ultimately overcome by the $J_{\rm ext}$-induced repulsion. 

These finite-size insights, reminiscent of other local second-order pairing mechanisms~\cite{kohn1965new,crepel2021new,crepel2022unconventional,crepel2024attractive}, shows that purely in-plane spin exchange may drive an effective pairing between spinons. 
In fact, the spin model Eq.~\ref{eq_spinmodel} shares many common features with the chiral tripod considered in Fig.~\ref{fig_smallclusters}c: a central site is coupled by a chiral in-plane exchange to other spins forming an outer ring, and spin within the outer rings interact through a SU(2) Heisenberg exchange. 
The intuition built from the finite clusters foretells the presence of spinon pairing, at least when $J' \ll J$, which is the case for the system considered (Fig.~\ref{fig_tworegimes}a).

\bibliography{spinonmeanfield}

\end{document}